\documentclass[3p,times,procedia]{elsarticle}
\usepackage{nupha_ecrc}

\volume{00}

\firstpage{1}

\journalname{Nuclear Physics A}

\runauth{}

\jid{nupha}

\jnltitlelogo{Nuclear Physics A}

\usepackage{amssymb}

\usepackage[figuresright]{rotating}

\begin{document}

\begin{frontmatter}



\dochead{XXVIIIth International Conference on Ultrarelativistic Nucleus-Nucleus Collisions\\ (Quark Matter 2019) \vspace{-0.1in}} 

\title{Chirality and Magnetic Field \vspace{-0.1in}}


\author[add1]{Defu Hou}
\author[add4]{Anping Huang}
\author[add4]{Jinfeng Liao \footnote{ presenter (liaoji@indiana.edu)}} 
\author[add2]{Shuzhe Shi}
\author[add3]{Hui Zhang }

\address[add1]{Institute of Particle Physics (IOPP) and Key Laboratory of Quark and Lepton Physics (MOE),  Central China Normal University, Wuhan 430079, China.} 
\address[add4]{Physics Department and Center for Exploration of Energy and Matter,
Indiana University, 2401 N Milo B. Sampson Lane, Bloomington, IN 47408, USA.} 
\address[add2]{Department of Physics, McGill University, 3600 University Street, Montreal, QC, H3A 2T8, Canada.} 
\address[add3]{Guangdong Provincial Key Laboratory of Nuclear Science, Institute of Quantum Matter, South China Normal University, Guangzhou 510006, China.\vspace{-0.3in}}

\begin{abstract}
We present a brief overview on recent developments of theory and phenomenology for novel many-body phenomena related to the chirality  and magnetic field, with an emphasis on their experimental implications and possible detection in relativistic nuclear collisions.  
\end{abstract}


\end{frontmatter}


\section{Introduction}
\vspace{-0.1in}
\label{}

In the past two decades or so, the heavy ion physics program has achieved great success by creating, measuring and understanding the quark-gluon plasma (QGP) at the Relativistic Heavy Ion Collider (RHIC) and the Large Hadron Collider (LHC). In particular, the QGP has been found as a nearly perfect fluid with its shear viscosity to entropy density ratio close to the quantum lower bound. This characterization has largely relied upon  thorough investigations of the {\em energy and momentum transport}  in these collisions and quantitatively contrasting comprehensive experimental measurements with viscous-hydrodynamic simulations.  It is only very recently that the community has started to scrutinize the behavior of another important degrees of freedom, namely the  {\em spin transport} in such a quantum fluid. 

\begin{figure}[htb!]
\begin{center}
\includegraphics[height=3.3cm]{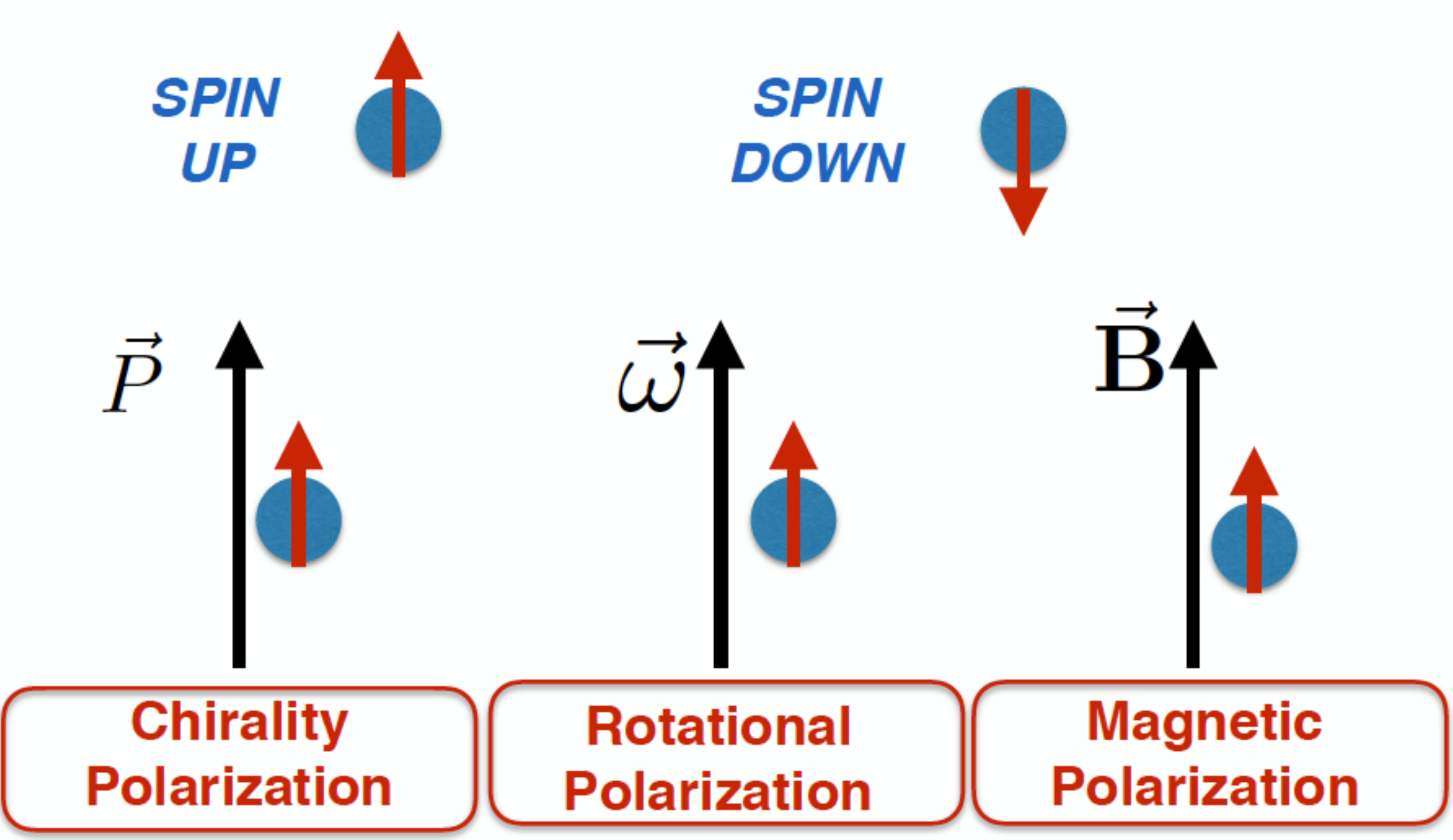}\hspace{0.4in}
\includegraphics[height=3.3cm]{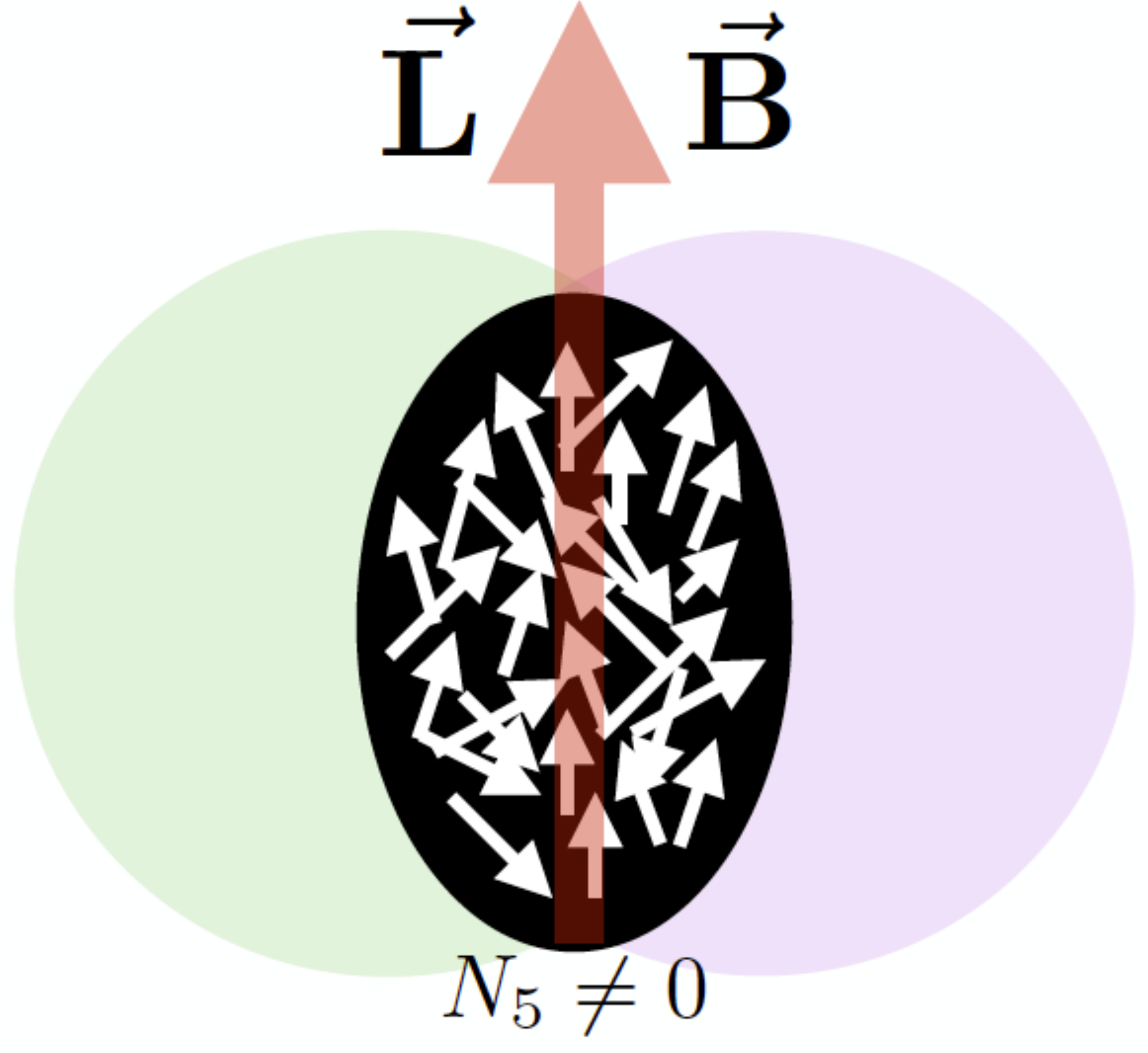}
\caption{(color online) \label{fig_spin}
 Illustration of fermion spin polarization due to chirality, vorticity and magnetic field (left) and illustration of the fireball created in heavy ion collisions as a ``spin fluid'' under the presence of macroscopic chirality imbalance as well as extremely strong vorticity and magnetic field  (right).   }\vspace{-0.2in}
\end{center}
\end{figure}

Spin is by nature a quantum degree of freedom. To be specific, let us focus the discussion on the fermion spin, i.e. that of the quarks in our QGP. The quark spin can point up or down along any particular direction specified by an external probe. There exist a number of interesting ways for ``playing'' with spin, notably through {\em chirality, vorticity and magnetic field} as illustrated in Fig.~\ref{fig_spin}(left). Each of these  ``handles'' could influence spins of microscopic particles (i.e. quarks in our consideration) by polarizing them: net chirality leads to a preference of the spins along or against (depending on sign of chiral charge) the momentum direction, vorticity enforces the spins to align more with the system's angular momentum, while magnetic field orients the spins to be preferably   along or against its direction (depending on sign of electric charge).    

What has attracted particular interests and research efforts is the investigation of {\em novel macroscopic phenomena} arising from interplay between microscopic spins with chirality, vorticity and magnetic field. Anomalous transport processes, such as the Chiral Magnetic Effect (CME) and Chiral Vortical Effect (CVE), have been enthusiastically studied across physics disciplines. Nontrivial influence on phase structures of matter has been identified. In the context of heavy ion collisions, the created fireball (in non-central collisions) inherits  a large angular momentum of the order $\sim 10^{4\sim 5} \hbar$ which induces strong vorticity field in the fluid. Due to the large positive charge and high speed of the initial ions, the fireball also experiences an extremely strong magnetic field of the order $\sim 10^{17\sim 18} \rm Gauss$. Furthermore a nonzero axial charge, i.e. imbalance between right-handed (RH) and left-handed (LH) quarks, is generally expected to occur due to topological fluctuations of the QCD gluon fields. Therefore the hot QCD fluid in relativistic nuclear collisions provides a unique subatomic material for understanding properties of matter under chirality, vorticity and magnetic field, as illustrated in Fig.~\ref{fig_spin}(right). 

This research topic has flourished in the last several years with exciting results coming from many fronts of theory, phenomenology and experiment, as reflected in a series of ``Chirality'' conferences~\cite{Chirality}. 
The present writeup, based on the talk ``Chirality and Magnetic Field'' delivered at the Quark Matter 2019 conference, will give a brief overview on recent developments of theory and phenomenology along this line of research,  highlighting a selective set of latest results. For the topic of vorticity, see contribution by X. Huang in this Proceedings. For experimental overview, see contribution by M. Lisa in this Proceedings. For more detailed discussions and sources of bibliography, see a number of excellent reviews in \cite{Bzdak:2019pkr,Kharzeev:2015znc,Li:2020dwr,Liu:2020ymh,Zhao:2019hta,Wang:2018ygc,Xu:2018fog,Hattori:2016emy,Fukushima:2018grm,Miransky:2015ava}.

\vspace{-0.1in}
\section{Chirality and chiral materials}
\vspace{-0.1in}

Quantum anomaly is a fundamental feature of chiral fermions. While the classical Lagrangian of massless fermions with gauge interactions has a global axial symmetry with a corresponding conserved axial current, such symmetry is spoiled by quantization in a specific way with a corresponding violation of axial current conservation. This provides a mechanism of axial charge (i.e. chirality imbalance) generation, for which new insight on its relation to Schwinger mechanism was obtained recently~\cite{Copinger:2018ftr}. The light flavor quarks in QCD are (approximately) massless and the non-Abelian chiral anomaly arising from their coupling to the gluon fields connects the chirality imbalance with gluon field topology: $N_5= N_R - N_L = 2 Q_w$ ,   where $N_5$ is the axial charge (per each flavor) quantifying the number difference between RH and LH quarks while the integer $Q_w$ is the topological winding number of gluon fields. This is to say,  every gluonic topological transition of nonzero $Q_w$ induces  a corresponding fluctuation in  the quark chirality imbalance $N_5$. An experimental measurement of the latter would be a unique and direct probe of the former. In addition, this relation bears a deep connection to mathematics, being essentially a special case of the Atiyah-Singer index theorem  for the Dirac operator in the instanton background fields. 

A rapidly emerging new frontier for the anomaly physics, is to understand its macroscopic implications for  {\em chiral materials}, which are many-body quantum systems that consist of  chiral fermions and could maintain a nonzero macroscopic chirality over long time scale. It turns out,  {\em the microscopic quantum anomaly}  manifests itself in these materials by inducing highly nontrivial   {\em macroscopic anomalous transport processes}  that are normally forbidden  but become possible (and necessary) in such P- and CP-odd environment. A notable example is the Chiral Magnetic Effect (CME), predicting the generation of an electric current in  chiral materials as response to an applied magnetic field: $\mathbf{J} =\sigma_5 \mathbf{B}$, where the conductivity $\sigma_5$ is directly proportional to chirality imbalance. The CME is a remarkable example as a new kind of quantum electricity that one may tentatively call ``magne-tricity''.  Enthusiastic efforts have been made for observing the CME in various physical systems. For example, in novel topological phases of condensed matter systems known as Dirac and Weyl semimetals the CME-induced transport has been measured via  observables like negative magnetoresistance~\cite{Li:2014bha}  (see review in e.g. \cite{Armitage:2017cjs}). New ideas based on CME  have also been actively explored across disciplines, such as quantum computing with chiral qubit \cite{Kharzeev:2019ceh,Chernodub:2019xjl} as reported by D. Kharzeev in this Proceedings. In addition to anomalous transport, there are also other   interesting  consequences  on properties of matter arising from interplay between microscopic spin and macroscopic probes such as new phase structures~\cite{Jiang:2016wvv,Zhang:2018ome}. 
Last but not least, notable progress has been and continues to be achieved in building the quantum transport theory for chiral materials~\cite{Gao:2018wmr,Gao:2019znl,Sheng:2018jwf,Weickgenannt:2019dks,Wang:2019moi,Huang:2018wdl,Li:2019qkf,Liu:2018xip,Mueller:2017arw,Mueller:2017lzw,Shi:2020qrx}, as reflected in the contributions to this Proceedings by J. Gao, Q. Wang, Z. Wang, N. Weickgenannt, S. Li, S. Shi, etc.

The quark-gluon plasma created in relativistic heavy ion collisions has offered the unique example of a subatomic chiral material with intrinsic relativistic  fermions. A key element at stake is the presence and fate of macroscopic chirality imbalance i.e. the axial charge $N_5$ in the hot QGP. In a typical collision, the fireball possesses considerable initial axial charge $N_5$ from random local and global topological fluctuations of strong gluon fields, albeit with equal probability to be positive or negative from event to event. This has been nicely demonstrated in recent  simulations performed in the initial-stage glasma framework~\cite{Mace:2016svc,Mace:2016shq,Lappi:2017skr} (--- see also early idea in \cite{Kharzeev:2001ev}).  Given a nonzero initial $N_5$, there is still the question of its subsequent relaxation  toward its vanishing equilibrium value. Both finite quark masses and gluonic topological fluctuations cause random flipping of individual quark chirality and  contribute to this relaxation rate. In the low temperature phase, a large dynamical mass for constituent quarks from spontaneous chiral symmetry breaking would quickly spoil any finite chirality. In the high temperature phase with restored chiral symmetry, the effect of  current quark masses (even for strange quarks) on axial charge relaxation turns out to be rather minimal as shown in recent analyses~\cite{Guo:2016nnq,Hou:2017szz,Lin:2018nxj}.  A realistic evaluation of both gluonic and mass contributions to axial charge relaxation~\cite{Hou:2017szz} suggests that the QGP would be able to maintain its finite chirality for considerable time during the dynamical evolution after a collision.

To wrap up the discussion thus far, the quark-gluon plasma as a chiral material provides the opportunity to observe the Chiral Magnetic Effect. Measuring this effect would open a  window for characterizing the intriguing topological fluctuations of QCD gluon fields. What's more, its detection would also  be an experimental evidence  for  the high-temperature restoration of QCD chiral symmetry. For the rest of this article, we will focus on the extreme magnetic field and the CME-induced signals in heavy ion collisions. 

\vspace{-0.1in}
\section{Extreme magnetic field}
\vspace{-0.1in}

Heavy ion collisions create an environment with extreme magnetic field originating from the fast-moving, high-charged nuclei. A simple estimate gives $|e\mathbf{B}|\sim \frac{\alpha_{EM} Z\gamma b}{R_A^2} \sim m_\pi^2$ at the center point between two colliding nuclei upon initial impact. This is among the strongest known magnetic field, even much larger than that of a magnetar. Given such magnetic field and a chiral QGP, one expects the CME to occur. The natural question then is: what would be its possible experimental signatures? For that matter, two crucial factors need to be understood: its azimuthal orientation as well as its time duration.   

\begin{figure}[htb!]
\begin{center}
\includegraphics[width=13cm]{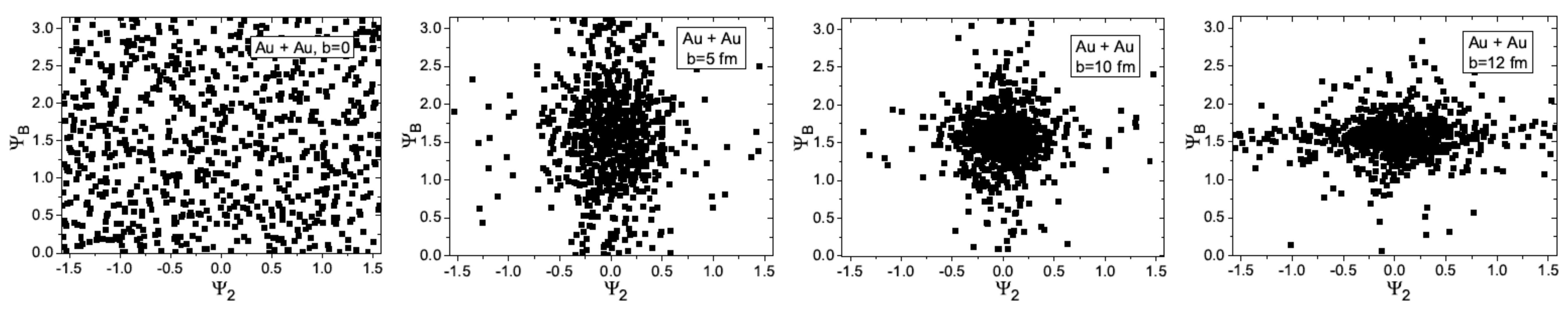} 
\caption{(color online) \label{fig_B1}
Scatterplot for $\mathbf{B}$ field azimuthal orientation $\Psi_B$ versus participant plane  $\Psi_2$   in $\sqrt{s_{NN}}=200\rm GeV$ AuAu collisions~\cite{Bloczynski:2012en}. See text for detailed discussions.} \vspace{-0.2in}
\end{center}
\end{figure}

While a simple picture based on optical collision geometry would suggest that the magnetic field points precisely along the out-of-plane direction (i.e. perpendicular to the reaction plane), this is not exactly true in reality. As first studied and emphasized in  \cite{Bloczynski:2012en}, strong fluctuations in the  initial conditions not only cause the matter geometry (as usually quantified by a series of harmonic participant planes) to vary considerably from event to event, but also bring significant fluctuations to the azimuthal orientation of the $\mathbf{B}$ field.  Fig.~\ref{fig_B1} (from \cite{Bloczynski:2012en}) shows a scatterplot of elliptic participant plane angle $\Psi_2$ and $\mathbf{B}$ field angle $\Psi_B$, both measured with respect to the reaction plane (RP), for various centrality of AuAu collisions at $200\rm GeV$. The $\Psi_B$ becomes more and more tightly correlated with RP from central to peripheral collisions, as $\Psi_B$ is dominated by spectators. The correlation between $\Psi_B$ and $\Psi_2$ becomes strong only in relatively  peripheral (but not too peripheral) collisions, and is generally weaker than that between $\Psi_B$ and $\Psi_{RP}$ for all centrality. 

\begin{figure}[htb!]
\begin{center}
\includegraphics[width=13cm]{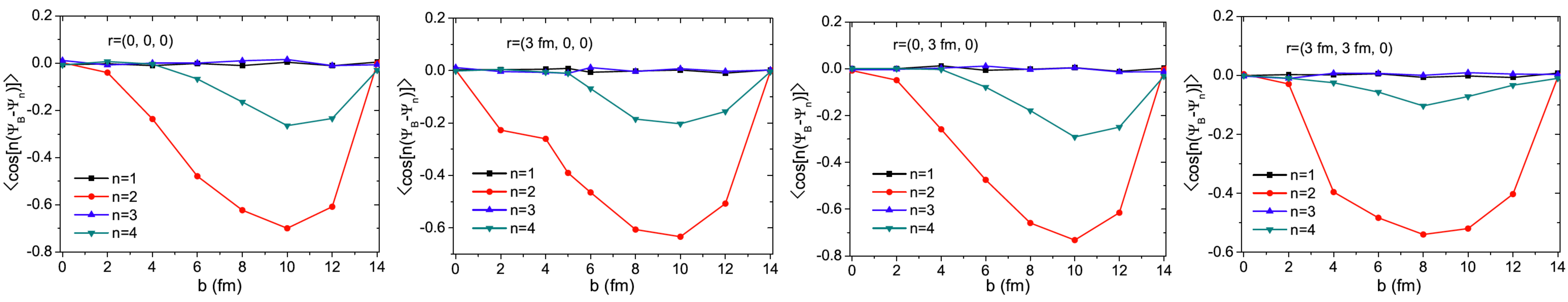} 
\caption{(color online)  \label{fig_B2}
Correlation factor $\langle \cos[n(\Psi_B - \Psi_n)] \rangle$ between $\mathbf{B}$ field  orientation $\Psi_B$ and fireball geometry via participant plane harmonics $\Psi_n$ ($n=1,2,3,4$)  in $\sqrt{s_{NN}}=200\rm GeV$ AuAu collisions~\cite{Bloczynski:2012en}. See text for detailed discussions. } \vspace{-0.2in}
\end{center}
\end{figure}

One can quantify the azimuthal correlation (or de-correlation) between magnetic field and matter geometry via the factor $\langle \cos[n(\Psi_B - \Psi_n)] \rangle$  for various harmonic planes, as shown in Fig.~\ref{fig_B2} (from \cite{Bloczynski:2012en}). The results clearly demonstrate that $\Psi_B$ is best correlated with elliptic plane (especially from mid-central to mid-peripheral region) while is {\em not} correlated with the triangular plane. Such important azimuthal fluctuations of the magnetic field turn out to be useful features for measurements, implying similar azimuthal correlation patterns of any $\mathbf{B}$-induced effect with the bulk matter geometry.  

\begin{figure}[htb!]
\begin{center}
\includegraphics[width=4cm,height=3.3cm]{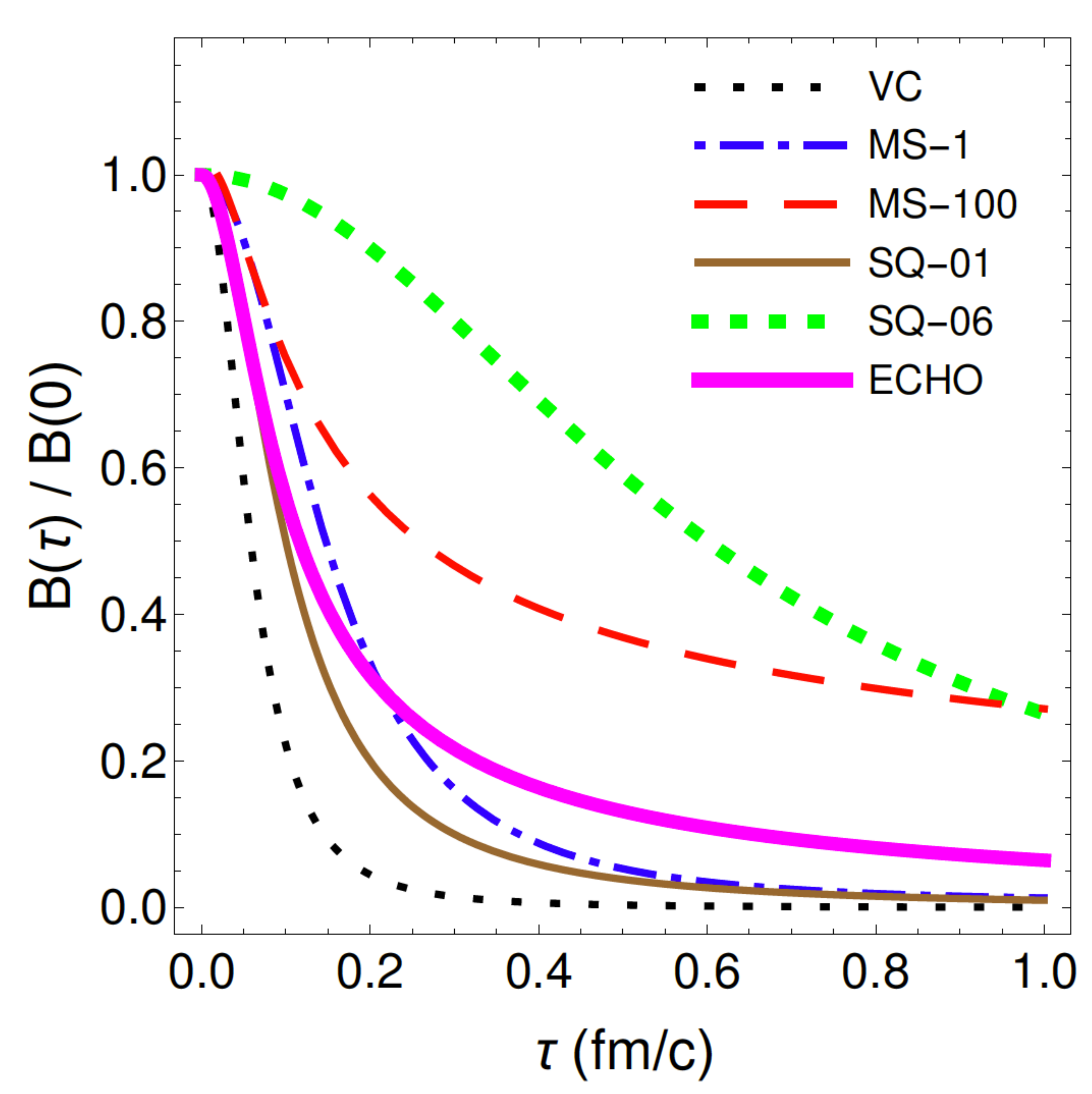}  \hspace{0.1in}
\includegraphics[width=4.5cm,height=3.3cm]{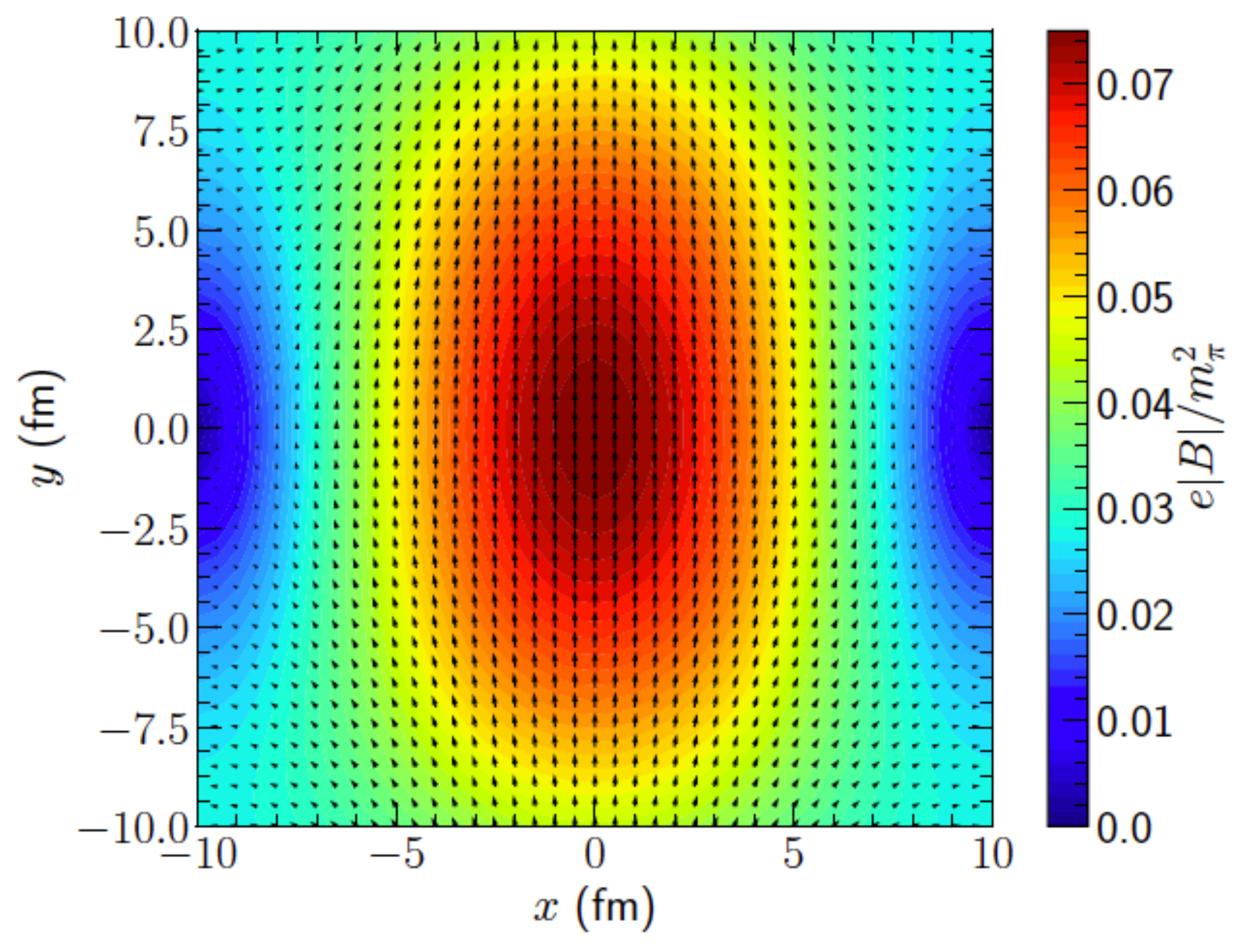}   \hspace{0.1in}
\includegraphics[width=4cm,height=3.3cm]{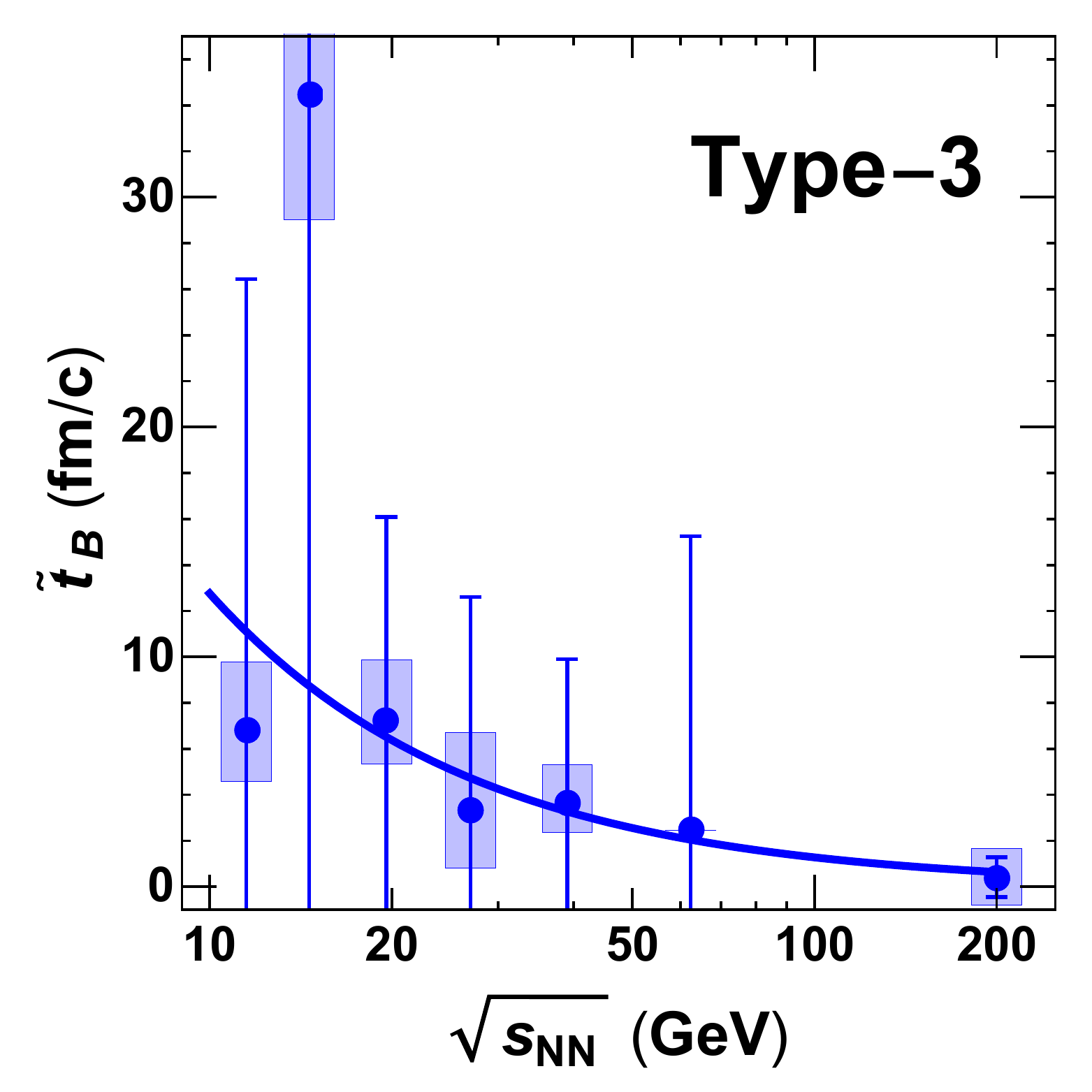}  \hspace{0.1in}
\caption{(color online)  \label{fig_Bt}
The left panel shows a compilation (from \cite{Huang:2017tsq}) of dynamical magnetic field time evolution from several calculations with varied assumptions and parameters about medium feedback effect. The middle panel shows a recent calculation~\cite{Gursoy:2018yai} of dynamical magnetic field distribution  on the transverse plane  after  evolving together with a hydrodynamic medium for a time of $\tau=1\rm fm/c$. The right panel shows an extraction of in-medium magnetic field lifetime versus collision energy $\sqrt{s_{NN}}$ across the RHIC beam energy scan range based on the measured global polarization difference between hyperons and anti-hyperons~\cite{Guo:2019joy}. 
} \vspace{-0.36in}
\end{center}
\end{figure}

While the initial magnetic field is strong, it rapidly decays over a short period of time, on the order $(R_A/\gamma)\sim 0.1\rm fm/c$, due to the departure of spectators down the collision pipeline. On the other hand, the created hot medium with quarks is an electric conductor and could potentially impede the decay of in-medium $\mathbf{B}$ field via the induction mechanism (i.e. Lenz's law). To figure out exactly how it evolves would require a realistic simulation of dynamical magnetic field along with the evolving medium itself. Many attempts with varied degrees of rigor and approximations were made~\cite{McLerran:2013hla,Tuchin:2015oka,Inghirami:2016iru,Inghirami:2019mkc,Gursoy:2018yai,Roy:2017yvg,Pu:2016ayh,Muller:2018ibh,Guo:2019joy,Guo:2019mgh}. The compilation in Fig.~\ref{fig_Bt} (left) shows wide-spread results from different calculations. It is clear from the plot that medium feedback could help elongate the lifespan of $\mathbf{B}$ field. It is also clear that the time dependence crucially depends on modeling details and  a final quantitative answer is not reached yet.  

One approach of describing the dynamical magnetic field is to develop magneto-hydrodynamic (MHD) framework~\cite{Denicol:2018rbw,Denicol:2019iyh,Siddique:2019gqh} and simulations~\cite{Inghirami:2016iru,Inghirami:2019mkc} for heavy ion collisions. Here the challenge is that the QGP may not be in an ideal MHD regime while going beyond that into resistive MHD is numerically very difficult. A less ambitious (albeit perhaps reasonably realistic) approach aims to solve the in-medium Maxwell's equations along with expanding conducting fluid while neglect the feedback of $\mathbf{B}$ field on the medium bulk evolution. Results from a recent attempt~\cite{Gursoy:2018yai} along this line is shown in Fig.~\ref{fig_Bt} (middle).   

Apart from CME, there are other interesting effects that can be directly induced by strong magnetic field and that could possibly be observed. For example, the $\mathbf{B}$ contributes an opposite polarization effect to the spins of particles and anti-particles (owing to their opposite magnetic moments). As magnetic field aligns with global angular momentum in a heavy ion collision, this may  provide a natural clue to  the puzzling observation of a larger global spin polarization of anti-hyperons than that of hyperons over a wide span of collision beam energy. Note the hadrons are produced at relatively late time in these collisions, so the observed polarization splitting could actually be used to help constrain/extract late time magnetic field~\cite{Muller:2018ibh,Guo:2019joy}, which in turn helps indirectly infer the 
$\mathbf{B}$ field lifetime. Such a quantitative analysis was performed in \cite{Guo:2019joy}, with one example result in Fig.~\ref{fig_Bt} (right). The magnetic field lifetime is estimated to be: $\tau_B \simeq \frac{(115 \pm 16) \,{\rm GeV \cdot fm /c}}{\sqrt{s_{NN}}}$.  This indicates a quite reasonable $\mathbf{B}$ field duration of $(0.5\sim 0.6)\ \rm fm/c$ at 200GeV collisions. 
At low-to-intermediate energy range the lifetime could be considerably long, and a novel mechanism based on charged fluid vortex for the generation of substantial late-time $\mathbf{B}$ field was proposed in \cite{Guo:2019mgh}. 

Despite current uncertain in its time evolution, it is obvious that the $\mathbf{B}$ is the strongest during the early moments after a collision. Therefore it would be sensible to look for  magnetic field effects  coming from such early stage. Two novel examples have attracted much interest recently. One is a predicted difference in the directed flow between particles and anti-particles, in particular for charm mesons~\cite{Inghirami:2019mkc,Gursoy:2018yai}. Both STAR and ALICE reported relevant measurements and a more detailed discussion can be found in the contribution by D. Kharzeev in this Proceedings. The other is the direct production of vector mesons and di-leptons from strong initial coherent nuclear electromagnetic fields~\cite{Abbas:2013oua,Khachatryan:2016qhq,Aaboud:2017bwk,Aaboud:2018eph,Adam:2018tdm}. In particular the intriguing measurements of transverse momentum broadening of produced di-leptons triggered a lot of discussions on their implications for possible in-medium magnetic field~\cite{Aaboud:2018eph,Adam:2018tdm,Zha:2018tlq,Zha:2018ywo,Klein:2018fmp}.

\vspace{-0.1in}
\section{Searching for Chiral Magnetic Effect}
\vspace{-0.1in}

\begin{figure}[htb!]
\begin{center}
\includegraphics[width=3.4cm,height=3.cm]{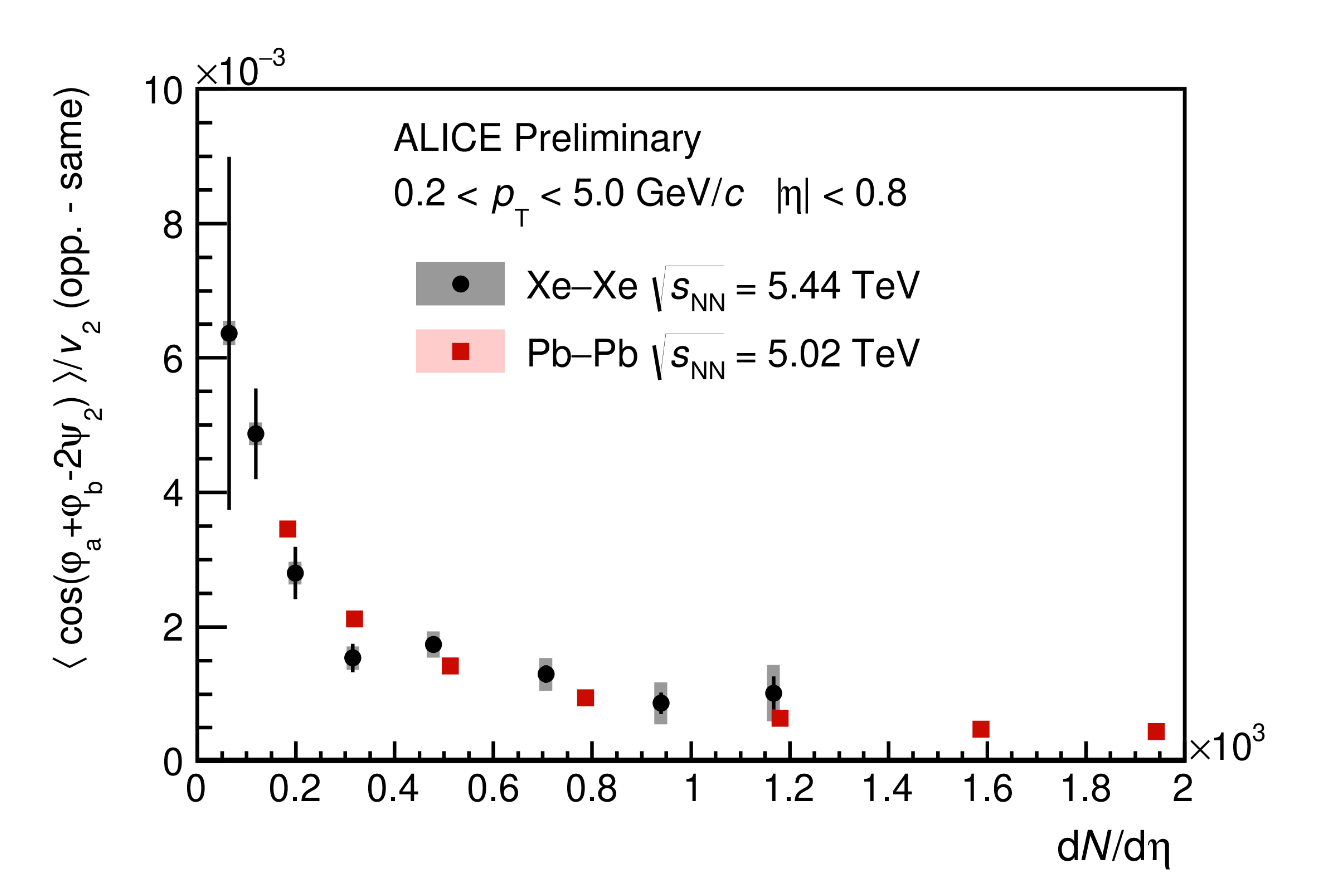}  \hspace{0.06in}
\includegraphics[width=3.4cm,height=3.cm]{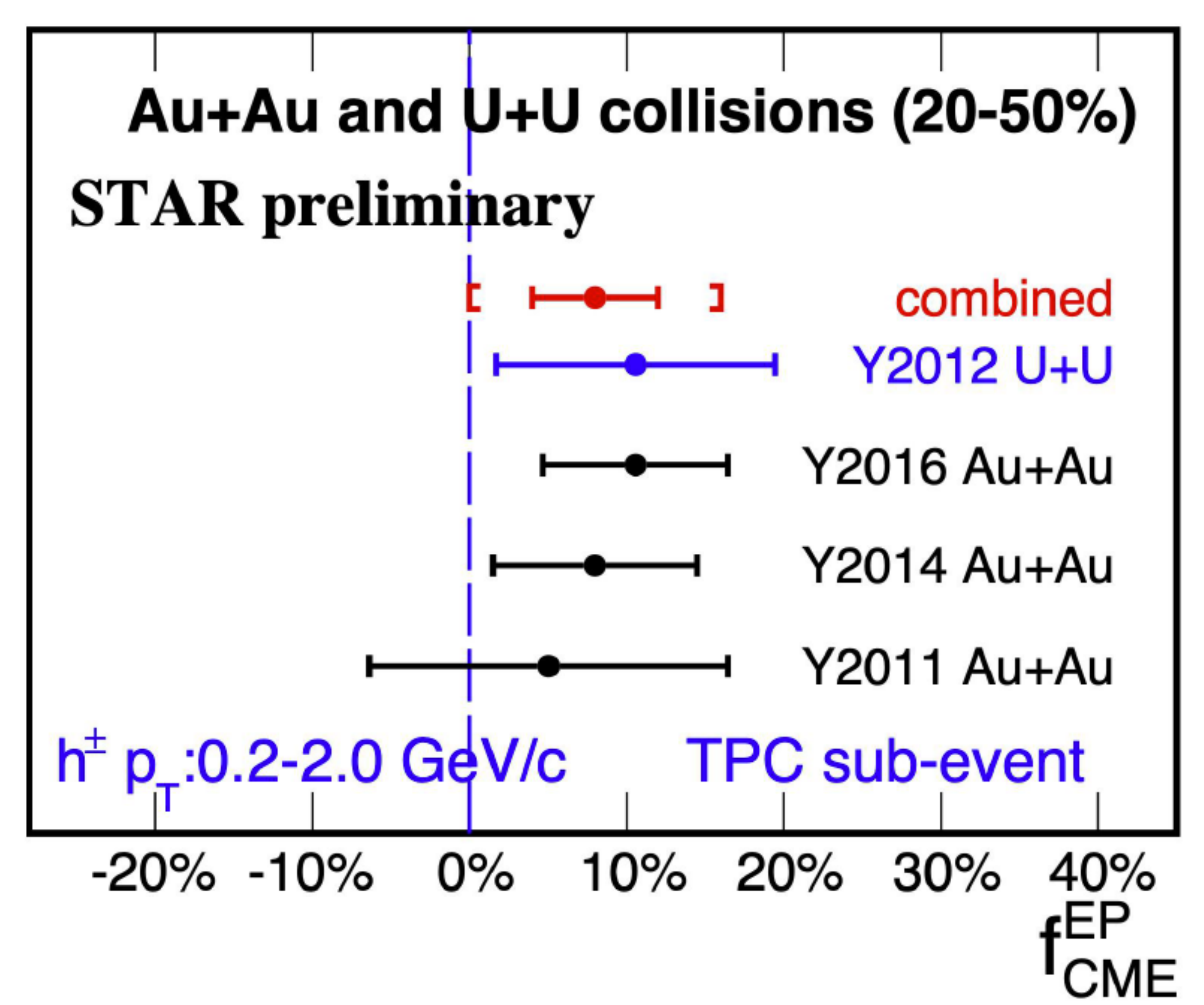}  \hspace{0.06in}
\includegraphics[width=3.4cm,height=3.cm]{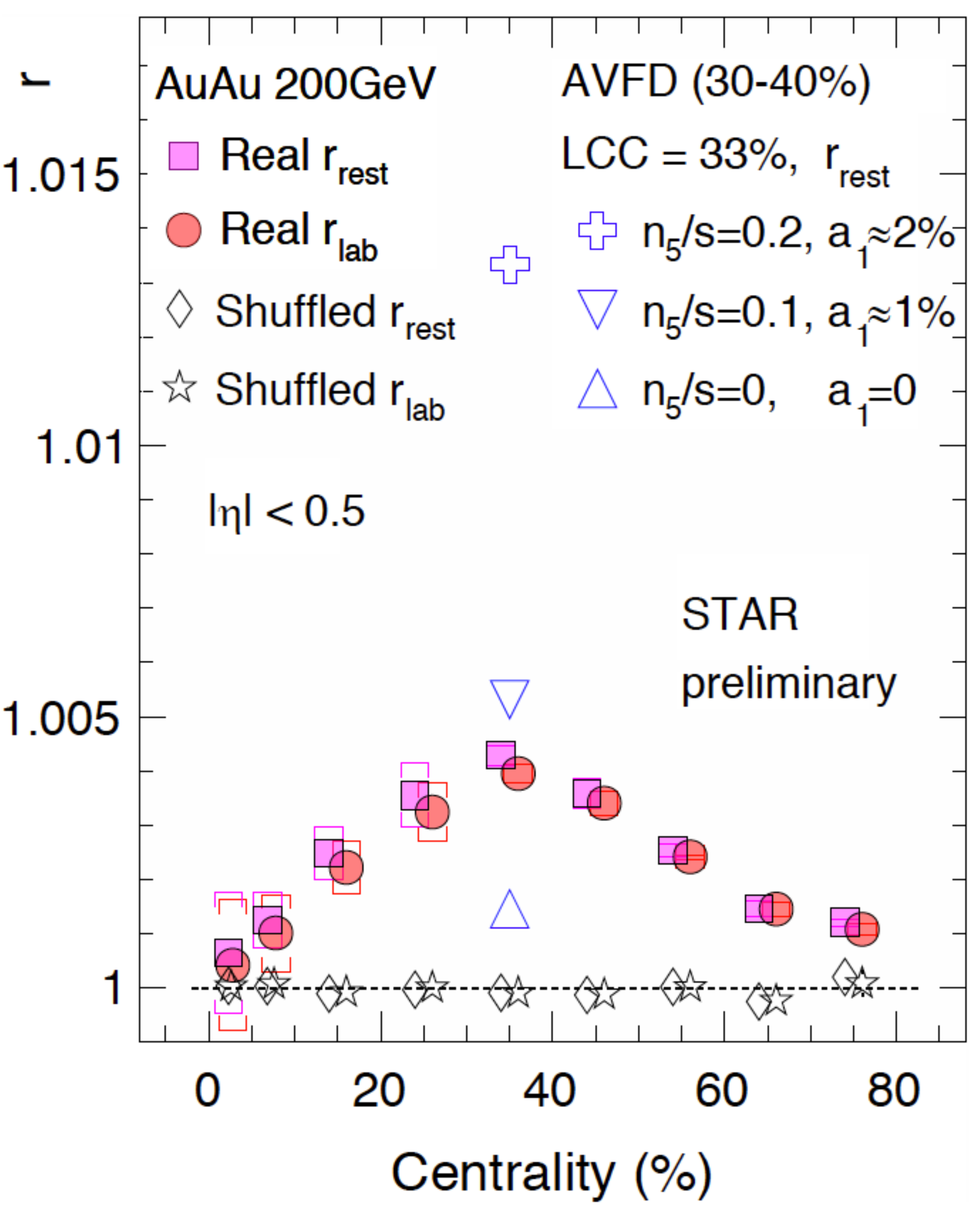}  \hspace{0.06in}
\includegraphics[width=3.4cm,height=3.cm]{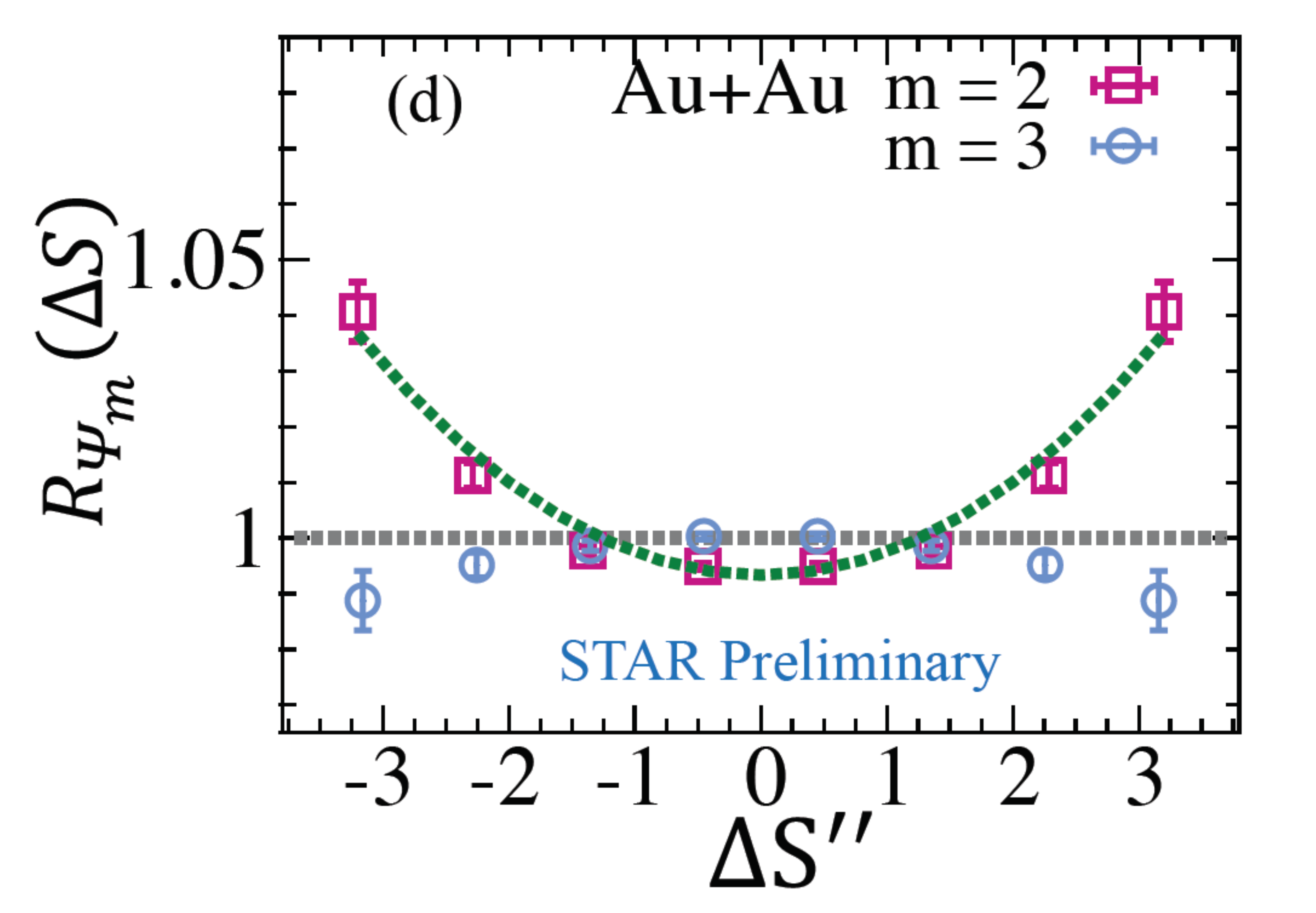} 
\caption{(color online)  \label{fig_exp}
A compilation of current CME measurement results  at RHIC and LHC. See details in the contributions by Z. Xu, by Y. Lin, by J. Zhao, by M. Weber and by S. Aziz in this Proceedings. These results would be challenging to interpret via a purely background scenario and appear to suggest a possibly detectable CME signal particularly in AuAu collisions at RHIC energy.} \vspace{-0.38in}
\end{center}
\end{figure}

As already discussed, potential discovery of CME in quark-gluon plasma  is of utmost significance. The CME-induced transport is expected to result in a dipole-like charge separation  along $\mathbf{B}$ field direction~\cite{Kharzeev:2004ey}, which could be measured by charge asymmetry in two-particle azimuthal correlations~\cite{Voloshin:2004vk}. 
Extensive experimental searches have been carried out over the past decade  to look for its traces by STAR at the Relativistic Heavy Ion Collider (RHIC) as well as by ALICE and CMS at the Large Hadron Collider (LHC). While lots of measurements, for a variety of colliding systems across a wide beam energy scan, have been accumulated so far  with encouraging hints, the interpretation of these data remains inconclusive due to significant background contamination --- see detailed discussions in e.g. \cite{Bzdak:2019pkr,Kharzeev:2015znc,Li:2020dwr,Zhao:2019hta,Bzdak:2012ia}. Several  methods have been developed to extract CME signals out of the overall correlations, with the latest experimental status summarized in Fig.~\ref{fig_exp} here for completeness. For a theorist's take on the implications of these measurements: (a) the total charge asymmetry correlations are dominated by backgrounds and the CME-contributed portion is likely only at the order of $(5\sim 10)\%$ level;  (b) despite its dominance, a scenario of pure backgrounds is unlikely to pan out in explaining existing data; (c) leaving the tricky systematic errors aside, from statistical point of view the current data are more consistent with the CME signal being nonzero than being zero.

To settle the presently controversial status,  a decisive  isobaric collision experiment was  carried out in the 2018 run at RHIC, with the dedicated physics goal of discovering the CME~\cite{Voloshin:2010ut,Skokov:2016yrj,Kharzeev:2019zgg}.  The basic idea is to contrast the CME-sensitive observables in two different colliding systems, the RuRu and the ZrZr, where the Ru and Zr are a pair of isobar nuclei with the same nucleon numbers ($A=96$) but different nuclear charges ($Z=44$ and $Z=40$ respectively).  The expectation is that the two systems will have the {\em same} background contributions while noticeably {\em different} CME signals   due to the difference in their  nuclear charge and thus magnetic field strength. This experiment offers the unique opportunity to detect CME in such collisions and currently the data analysis is actively underway\cite{Adam:2019fbq}.

On the theory side, a precise and realistic characterization of the CME signals and backgrounds is critically needed. To achieve this, a new tool, EBE-AVFD (Event-By-Event Anomalous-Viscous Fluid Dynamics)~\cite{Shi:2019wzi,Shi:2017cpu,Jiang:2016wve}, has been recently developed through the CME Working Group effort within the BEST Collaboration.  This is a state-of-the-art hydrodynamic framework implementing CME transport current in dynamically evolving bulk viscous fluid, incorporating late-time hadron cascade stage and taking into account the known major backgrounds. It is now widely used for developing and testing CME-motivated new observables~\cite{Magdy:2017yje,Tang:2019pbl}. (We also note the efforts  in simulating CME based on transport models~\cite{Deng:2018dut,Sun:2018idn,Zhao:2019crj}.)  

\begin{figure}[htb!]
\begin{center}
\includegraphics[width=3.3cm,height=3.2cm]{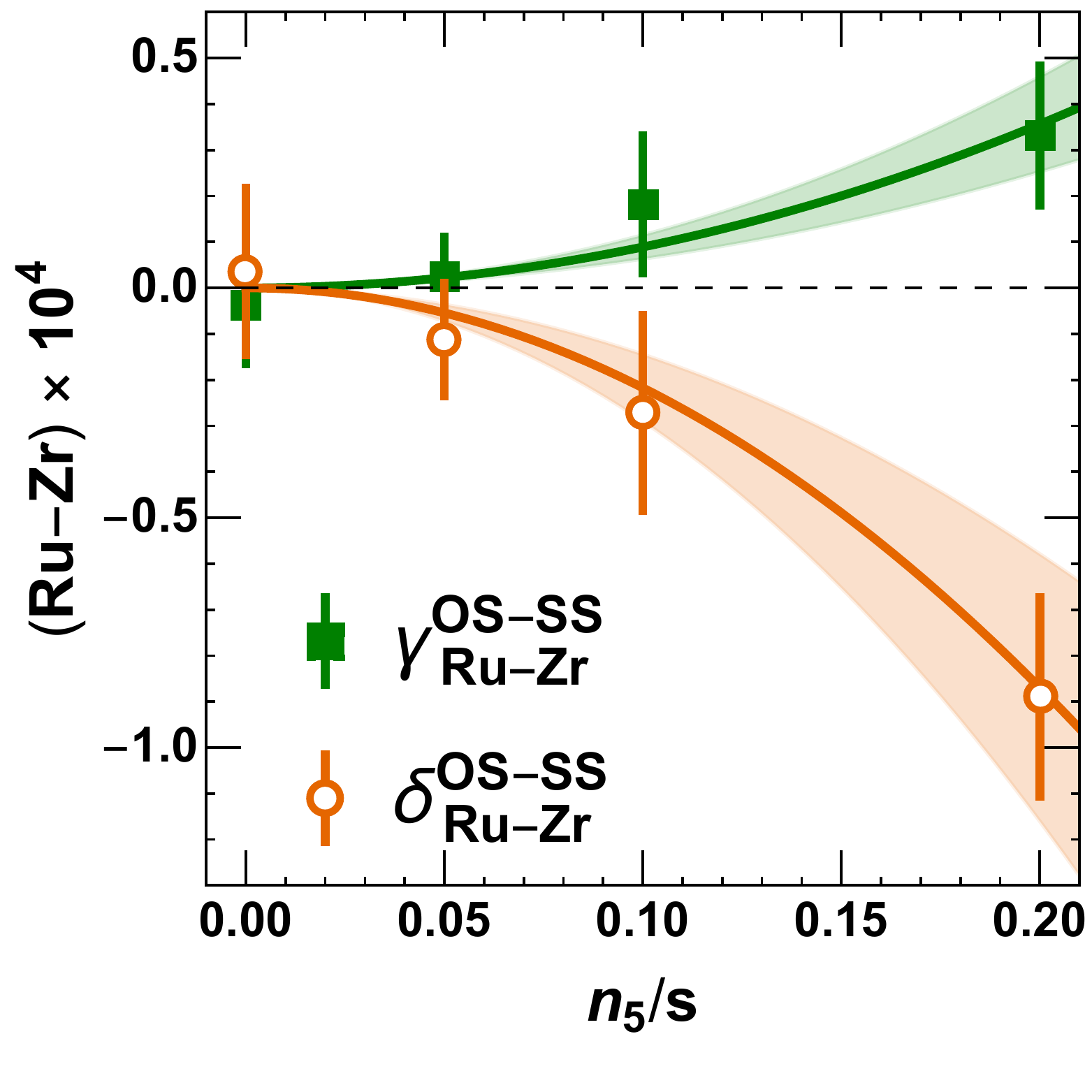} \hspace{0.1in}
\includegraphics[width=3.3cm,height=3.2cm]{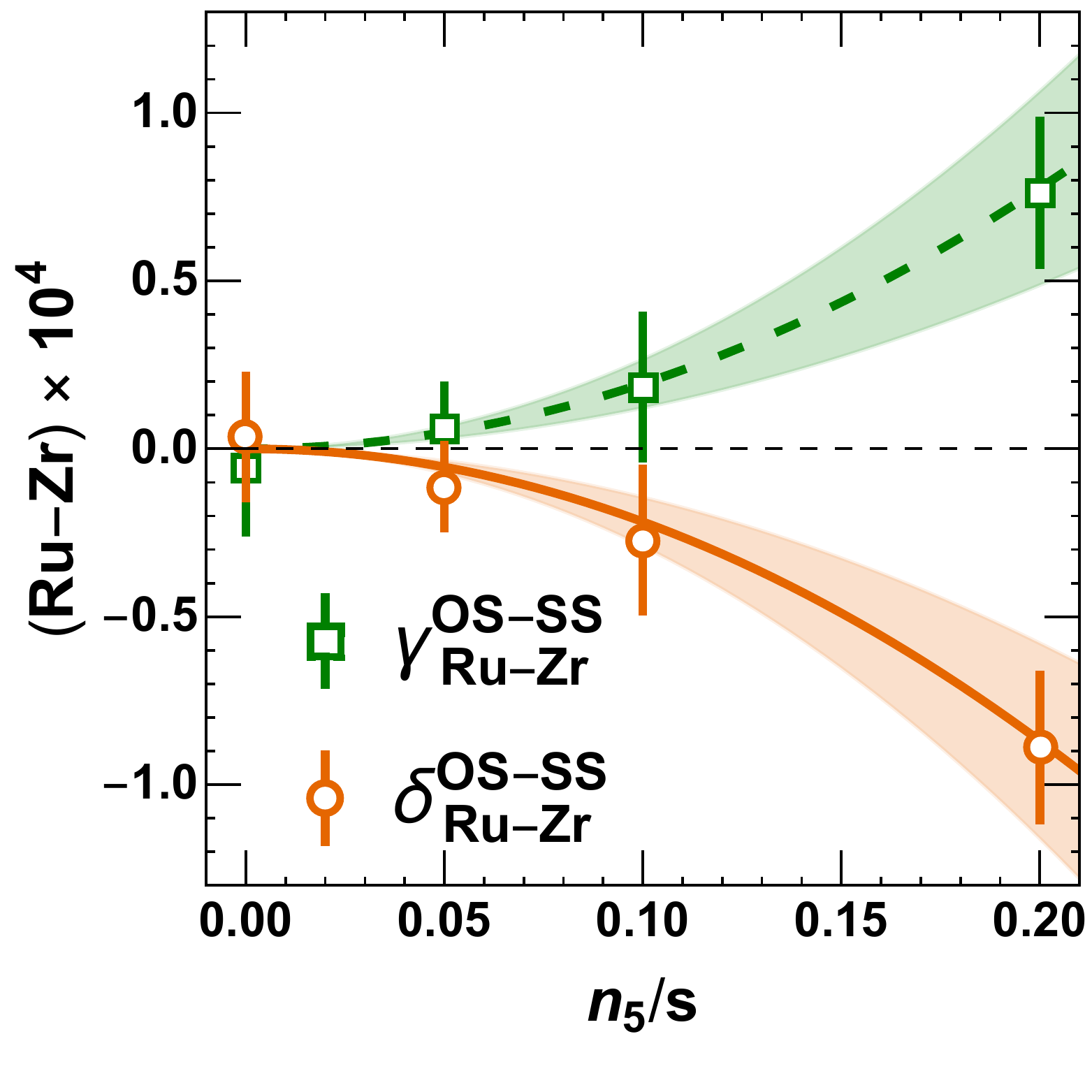} \hspace{0.1in}
\includegraphics[width=3.3cm,height=3.2cm]{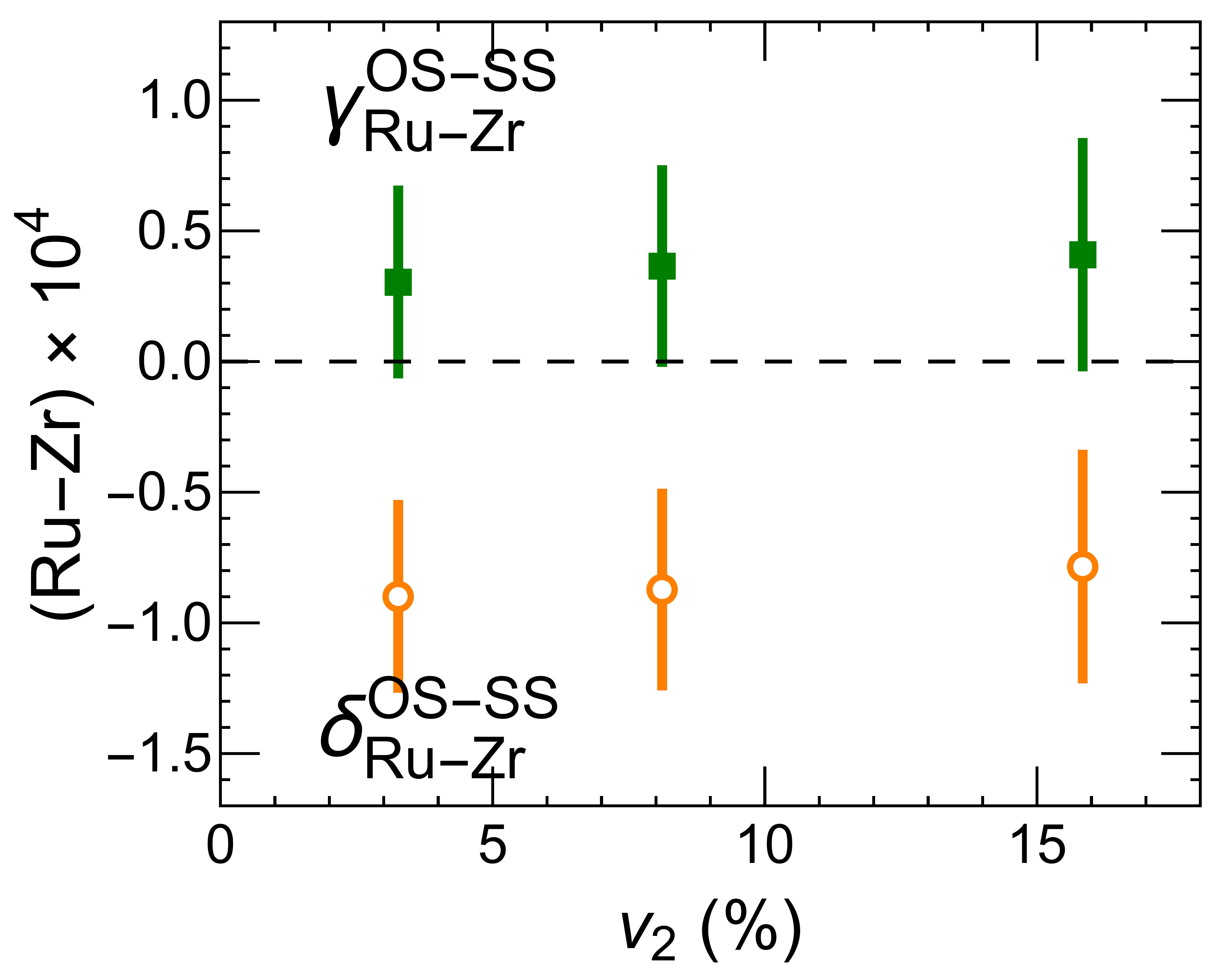} \hspace{0.1in}
\includegraphics[width=3.3cm,height=3.2cm]{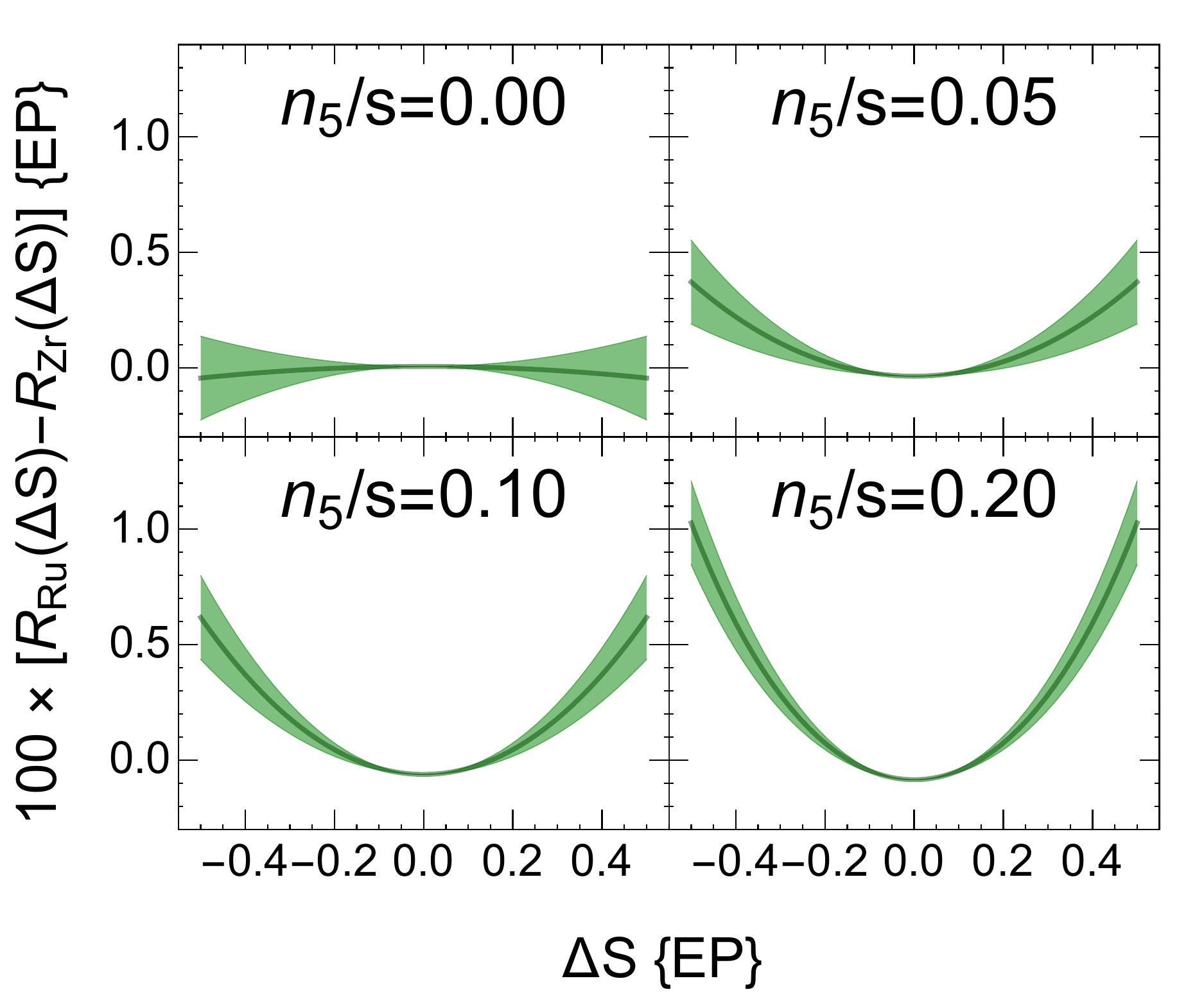}
\caption{\label{fig_EP} (color online)  \label{fig_isobar}
EBE-AVFD (Event-By-Event Anomalous-Viscous Fluid Dynamics)~\cite{Shi:2019wzi,Shi:2017cpu,Jiang:2016wve} predictions for experimental observables in isobaric collisions. Detailed discussions of the framework, observables and results are presented in ~\cite{Shi:2019wzi}.} \vspace{-0.27in}
\end{center}
\end{figure}

Equipped with this versatile tool, we are poised to make quantitative {\em predictions} for measurements to be reported in near future. In~\cite{Shi:2019wzi}, an optimal isobar comparison strategy is proposed and a set of detailed predictions for observables in these collisions were reported, with key results showcased here in Fig.~\ref{fig_isobar}.  

\vspace{-0.15in}
\section{Summary and outlook}
\vspace{-0.1in}

To summarize in one sentence: the Chiral Magnetic Effect in quark-gluon plasma embodies physics of gluon topology, quark chirality and quantum anomaly, whose observation in heavy ion collisions would be a fundamental discovery. As for outlook, it is simply an exciting time to have our fingers crossed and breaths held, counting down the {\em months} to come before  the anticipated release of isobaric collision measurements. 




This work is supported in part by the NSFC Grants No. 11735007 and No. 11875178, by the NSF Grant No. PHY-1913729 and by the U.S. Department of Energy, Office of Science, Office of Nuclear Physics, within the framework of the Beam Energy Scan Theory (BEST) Topical Collaboration. 
\vspace{-0.15in}

\bibliographystyle{elsarticle-num}
\bibliography{Chirality.bib}







\end{document}